\documentclass{article}
\usepackage[utf8]{inputenc} 
\usepackage[T1]{fontenc}    
\usepackage{hyperref}       
\usepackage{url}            
\usepackage{booktabs}       
\usepackage{amsfonts}       
\usepackage{nicefrac}       
\usepackage{microtype}      
\usepackage{graphicx}       

\usepackage[left=1.2in, right=1.2in, top=1in, bottom=1in]{geometry}
\usepackage[style=nature]{biblatex}
\usepackage[labelfont=bf,textfont=md]{caption}
\usepackage[table]{xcolor}
\usepackage{algorithm, algpseudocode}
\usepackage{authblk}
\usepackage{chemformula}
\newcommand{\siref}[1]{S#1} 

\usepackage{tabularx}
\usepackage{ragged2e} 
\newcolumntype{C}{>{\Centering\arraybackslash}X}
\usepackage[draft]{changes}

\title{Validating Graph Neural Network Feature Attribution Through Local Perturbations: A Study on CoMoFeNiCu Decomposition Enthalpy}
\title{Validating Graph Neural Network Feature Attribution Through Local Perturbations: A Study on CoMoFeNiCu Decomposition Enthalpy}

\author[1]{Miguel Tenorio}
\author[1]{James Chapman}
\affil[1]{Department of Mechanical Engineering, Boston University, Boston, MA 02215, USA}

\begin{document}                           
\maketitle


\begin{abstract}
    Graph neural networks (GNNs) show promise for predicting materials properties, but interpretability remains a challenge, particularly understanding whether attributed feature importance reflects physical reality or statistical artifacts. Here, we present a methods validation study that deliberately overfits a physics-aware GNN ensemble to probe the robustness of feature-ranking approaches. Using the CoMoFeNiCu high-entropy alloy system with three stoichiometries, we train GNNs to predict decomposition enthalpies from DFT-calculated structures while explicitly constraining output weights to encode feature importance. By design, our dataset is limited (approximately 900 binary and 4,500 ternary decomposition pathways) to control compositional complexity and enable precise interrogation of local chemistry effects. We validate that feature rankings capture meaningful local-environment information through a series of atomic swap perturbations: perturbing individual atomic sites and re-ranking features reveals spatial decay consistent with the GNN's interaction cutoff, confirming that weights encode neighbor-shell effects rather than global stoichiometric artifacts alone. Results demonstrate that feature attribution methods can remain physically interpretable even under overfitting, provided validation is rigorous. This work establishes a foundation for trustworthy interpretability in materials GNNs and sets the stage for scaling to multiple chemistries and testing generalization across compositional spaces.
\end{abstract}  
\section*{Introduction}

Graph neural networks (GNNs) have emerged as powerful tools for predicting materials properties from atomic structure, leveraging their ability to learn localized, permutation-invariant interactions through message passing and graph convolution operations \cite{gilmer2017neural, gori2005new}. Their success has motivated rapid adoption across computational materials science, from crystal property prediction to interatomic potential development \cite{xie2018crystal, chen_graph_2019, schutt2017schnet,reiser_graph_2022, shi_review_2024}. However, a critical gap remains: interpretability. While GNNs produce accurate predictions, understanding why a model makes a particular prediction, and crucially, whether attributed importance reflects genuine physical insight or merely statistical artifacts remains an open problem. This distinction matters profoundly: a feature-attribution method that conflates correlation with causation may offer false confidence in mechanistic understanding, potentially misleading materials design \cite{jiang2023graph}.

Recent advances in explainable artificial intelligence (XAI) have introduced feature-attribution techniques to assign importance weights to individual input features based on their contribution to model output \cite{nazir2025survey, bugueno2024graph}. These methods are computationally efficient and model-agnostic, making them attractive for GNN interpretation \cite{ying2019gnnexplainer}. However, they face a fundamental validation challenge where in real materials datasets, we rarely know ground truth. We cannot definitively separate whether high attribution scores reflect learned physical relationships or statistical overfitting. This problem is especially acute when: training data is limited (as is typical in high-throughput DFT studies), global composition dominates predictions, masking local chemistry effects, or when models are validated only on test sets drawn from the same statistical distribution as training data \cite{xu2023small, zhou2025machine, madras2019detecting,tenorio2026out}.

The adoption of XAI methods in computational materials science has accelerated rapidly, with practitioners increasingly using feature rankings and saliency maps to extract chemical insights from trained models \cite{kotobi2023integrating, wang_machine_2020, zhong2022explainable,braham2025generating}. Yet most studies in materials science do not rigorously validate whether these interpretations are physically meaningful or merely statistical artifacts \cite{li2025diverse}. The absence of controlled validation is troubling: if a model learns spurious correlations (ex: Co-richness as a proxy for some other property), feature attribution will faithfully report these correlations as ``important,'' potentially codifying false mechanistic understanding into design rules. Moreover, the challenge intensifies under realistic conditions,limited data, high-dimensional feature spaces, and compositional complexity, where overfitting is common and statistical fitting and physical insight become difficult to disentangle \cite{xu2023small, schmidt_recent_2019}. Bridging this gap requires deliberate experimental design: datasets engineered to isolate specific physics, models trained under known statistical conditions, and validation methods that directly probe whether learned features encode the mechanisms we believe they do.

To address this gap, we present a methods validation study that deliberately overfits a GNN on a chemically homogeneous, stoichiometrically diverse dataset, then rigorously tests whether feature rankings capture local atomic environments through targeted atomic perturbations. Our core hypothesis is that if feature attribution reflects genuine local chemistry, then perturbing individual atoms should produce localized changes in feature scores, with magnitude decaying over the model's interaction range. We focus on the CoMoFeNiCu multi principle element alloy (MPEA) system across three stoichiometries (varying only Co content while holding Fe, Ni, and Cu fixed), deliberately eliminating chemical complexity that could introduce spurious statistical artifacts \cite{xie2019highly, saidi2021optimization}. This design, single chemistry and multi-stoichiometry, is unconventional for a discovery study but ideal for methods validation: it isolates compositional variation from structural variation and ensures that any feature-ranking insights derive from local-environment changes rather than emergent chemistry effects \cite{miracle2017critical, otto2013relative}.

\begin{figure}[h!]
    \centering
    \includegraphics[width=1.0\textwidth]{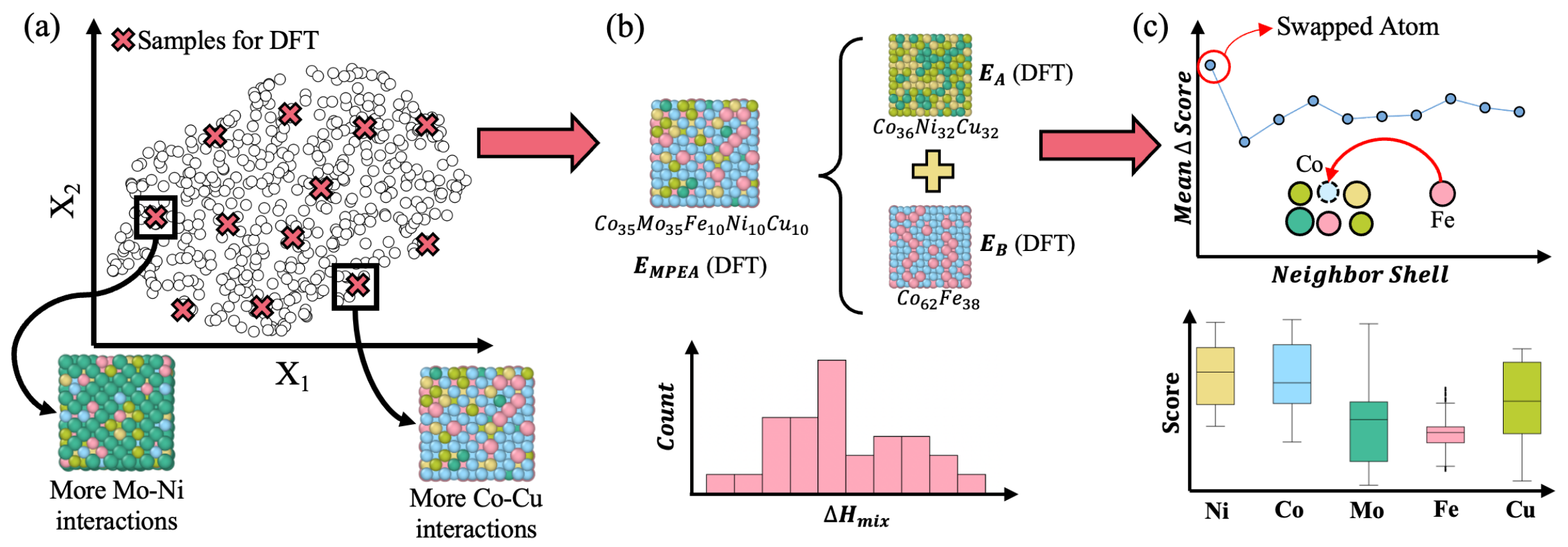}
    \caption{Three-stage computational framework for structure–property mapping of MPEAs. (a) An unsupervised GNN encoder projects randomly generated MPEA structures onto a latent space, where structurally distinct configurations, differing in local chemical ordering (e.g., Mo–Ni versus Co–Cu enrichment), are selected for DFT calculations. (b) Mixing enthalpies are computed by constructing decomposition pathways from DFT total energies of the MPEA and its constituent endpoint phases. (c) A supervised ML model trained on DFT labels to predict the target property, and ensemble-based feature ranking quantifies the relative contribution of each feature, like the contribution of each element}
    \label{fig:intro}
\end{figure}

Our approach unfolds in three stages. First, we use an unsupervised GNN encoder to project randomly generated MPEA structures onto a low-dimensional latent space, from which we sample strategically to balance the compositional-structural space before expensive DFT calculations. Second, we train a supervised GNN model (deliberately overfitting on our limited dataset) to predict mixing enthalpy from decomposition pathways. Third, we validate that feature rankings encode local chemistry by performing atomic swaps and re-ranking features across multiple stoichiometries, demonstrating spatial decay consistent with the model's message-passing range.

This work is intentionally scoped as a proof-of-concept for attribution robustness with a controlled dataset and training paradigm. Here, we ask a crucial question: can feature-attribution methods remain physically meaningful even when the statistical fit is imperfect? By answering  this question, we can establish a foundation for trustworthy GNN interpretability in materials science. The remainder of this paper is organized as follows. We begin by describing our dataset construction, DFT methodology, and graph formulation. We follow that with a discussion about our latent-space analysis, DFT-computed mixing enthalpies, ensemble predictions, and most importantly, the feature-ranking validation via atomic perturbations. We conclude with a discussion of implications and potential next steps for this work.

\section*{Methods}
\subsection*{Dataset Creation}

The structures used in this study were generated through a Python workflow utilizing the Atomic Simulation Environment (ASE) package \cite{HjorthLarsen2017TheAS}. An illustration of this structure-generation approach is shown in \textbf{Figure \siref{1}}. Three stoichiometries of the CoMoFeNiCu system were investigated, expressed as \textit{\(Co_xMo_{70-x}Fe_{10}Ni_{10}Cu_{10}\)}, with x taking values of 15, 35, and 55. For each stoichiometry, 500 randomly distributed MPEA configurations were generated.

\begin{table}[h!]
    \centering
    \caption{Summary of the dataset partitioning scheme, where $P_1$ and $P_2$ comprise binary and ternary decomposition pathways, respectively.}
    \label{Table:0}
    \begin{tabular}{|c c c c c|}
        \hline 
        \rowcolor[HTML]{99DDFF}
        \textbf{Pathway} & \textbf{Total Graphs} & \textbf{Training} & \textbf{validation} & \textbf{Testing}\\
        \hline \hline
        \(P_1\) & 877 & 631 & 70 & 176\\
        \(P_2\) & 4442 & 1599 & 177 & 445\\
        \hline
    \end{tabular}
\end{table}

Because the focus of this study centers on decomposition pathways rather than the MPEA alone, corresponding structures for the decomposition products were also constructed. Two types of decomposition pathways are considered: binary decomposition (\(P_1\)) and ternary decomposition (\(P_2\)) pathways. In a \(P_1\) pathway, the parent MPEA decomposes into two product phases, whereas in a \(P_2\) pathway it decomposes into three. In both cases, the product phases are required to preserve the stoichiometric ratios inherited from the parent MPEA, while their atom counts are adjusted only according to the symmetry constraints of their respective crystal structures. All MPEAs and alloy phases were modeled in the face-centered cubic (FCC) structure and contained 864 atoms, whereas certain single-element phases adopted a body-centered cubic (BCC) structure and contained 686 atoms. \textbf{Table \ref{Table:0}} summarizes the dataset breakdown into training, validation, and testing sets for both bathway types. It is worth noting that while the \(P_2\) dataset contains 4442 graphs in total, of which a randomly selected subset of 2221 was used due to hardware limitations, with the remaining 2221 graphs excluded from model development.

\subsection*{Electronic Structure Theory Details}

First-principles calculations were performed using the Vienna Ab initio Simulation Package (VASP) within the framework of spin-polarized DFT, using the Perdew–Burke–Ernzerhof revised for solids (PBEsol) functional under the generalized gradient approximation (GGA) \cite{kresse1999ultrasoft,kresse1996efficient,perdew1996generalized}. Electron–ion interactions were described using the projector augmented-wave (PAW) method \cite{Blochl1994}. A plane-wave kinetic energy cutoff of 500 eV was used throughout and pin polarization was explicitly included to capture the magnetic character of Fe- and Co-containing alloys. Geometry optimization (GO) was performed using the conjugate gradient algorithm with full relaxation of ionic positions, cell shape, and cell volume. Electronic convergence was enforced to within \(1.0^{-4}\) eV, and ionic relaxation was terminated when all forces fell below 0.1 $\frac{eV}{\text{\AA}}$. Gaussian smearing with a smearing width of \(\sigma = 0.1\) was used. Real-space projection was employed to reduce computational cost for the large supercells considered (864 atoms). The fully relaxed total energies were subsequently used to evaluate \(\Delta H_{mix}\) as described in the next section.

\subsection*{Phase Stability in MPEAs}
\label{subsec:phase}

To begin evaluating phase stability, we define the Gibbs free energy of mixing as follows:

\begin{equation}
    \Delta G_{mix} = \Delta H_{mix} - T\Delta S
\end{equation}

where \(\bigtriangleup H_{mix}\) represents the enthalpy of mixing, \(\bigtriangleup S\) is the entropy of the system, and \(T\) is the thermodynamic temperature. A negative value of \(\bigtriangleup G_{mix}\) indicates that the system energetically favors remaining an SPSS, whereas a positive value implies a preference for phase separation out of the SPSS. In this work, however, we are interested only in the $\Delta H_{mix}$ due to its intrinsic dependence on the DFT ground-state energy.

To compute \(\bigtriangleup H_{mix}\), it is necessary to construct decomposition pathways that model the MPEA breaking down into its constituent phases. These pathways are categorized by the number of resulting phases, for example, binary, ternary, or higher-order decompositions. The formulation we use for obtaining  \(\bigtriangleup H_{mix}\) is as follows:

\begin{equation}
    \Delta H_{mix} = E_{MPEA} - \sum_{i}^{N} E_{i}
\end{equation}

Where $i$ indexes the various $N$ decomposition structures possible for a specific decomposition pathway, and $E$ is a ground-state energy obtained from DFT. From these results: If \(\bigtriangleup H_{mix} \le 0\) the MPEA is stable at finite temperatures even without entropy contributions. If \(\bigtriangleup H_{mix} > 0\), the MPEA may become stable at finite temperatures due to entropy effects. Moreover, at \(T=0\), \(\bigtriangleup G_{mix}\) reduces to \(\bigtriangleup H_{mix}\). Within the scope of this study \(\bigtriangleup H_{mix}\) can therefore be interpreted as the likelihood that an MPEA remains a single-phase solid solution.

\subsection*{Graph Construction and Embedding for DFT Sampling}
\label{subsec:alignn}

We leverage a line graph formulation that consists of representing a single structure as two separate graphs: an atomic graph \(G\) where the atoms are represented as nodes and the edges represent the 2nd body interactions (bonds); and a line graph \(L(G)\) where the nodes represent the bonds and the edges the bond 3rd body interactions (bond angles). ASE and PyTorch Geometric \cite{fey2019fast} are used to generate this graph representation.

\begin{table}[h!]
    \centering
    \caption{ \label{Table:1} Model parameters for unsupervised graph embedding.}
    \begin{tabular}{|c c|}
        \hline 
        \rowcolor[HTML]{99DDFF}
        \textbf{Parameter} & \textbf{Value}\\
        \hline \hline
        Number of interaction layers & 6\\
        Radial Bessel basis cutoff  & 3.5 \AA \\
        Gaussian basis range & (-1,1) \\
        Number of channels & 100\\
        \hline
    \end{tabular}
\end{table}

To start the convolution or message-passing operations, all the scalar features (atom type, bonds, and bond angles) were converted to vector features. Following the work on \cite{aroboto2023universal}, we transformed the atom type using a learnable embedding matrix \cite{paszke2019pytorch}, expanded the bonds into the Radial Bessel basis \cite{gasteiger2020directional}, and lastly expanded the bond angles into the Gaussian basis \cite{schutt2017schnet}. It should be noted that the embedding is an unsupervised framework with no training. We used edge-gated graph convolutions for message-passing, as described in ALIGNN and \cite{aroboto2023universal}. By nature of the convolution operations, all the embedding vectors for all features during the interaction layers are the same length. All model parameters are summarized in \textbf{Table \ref{Table:1}}.

Since the original pool of candidate MPEA structures is too large to subject all of them to computationally intensive DFT calculations, down selecting structures from this pool becomes a necessary step. During sampling, it is essential to ensure that the selected structures adequately span the latent space. The output of the GNN embedding is a high-dimensional manifold with length equal to that of the initial layer. Although sampling could, in principle, be performed directly in this high-dimensional space, we instead exploit the fact that the GNN embedding encodes physically meaningful information about the graphs and the structures they represent. To enable both sampling and a visual interpretation of how the graphs populate the material space, we project the embedding onto a two-dimensional latent space using UMAP \cite{mcinnes2018umap} and perform sampling within this reduced representation. Specifically, we employ GRID sampling to obtain a comparable number of structures for each stoichiometry \cite{botu2017machine}: the two-dimensional latent space is partitioned into a regular grid of bins, and a single graph is randomly selected from every non-empty bin. This ensures that the sampled set reflects the full spread of the embedding rather than over-representing densely populated regions.

\subsection*{Graph Construction and embedding for Property Mapping}

We apply this same pipeline (latent-space graph embedding \(\rightarrow\) UMAP projection \(\rightarrow\) GRID sampling) to partition the data into training, validation, and testing sets for the ML models. In addition, we devise a way to encode the decomposition pathway directly into the graph formulation, which we call the Reaction Line Graph (RLG) representation. A generalized form of RLG is as follows:

\[G_{P} = \bigcup_{i}^{S} G_{i}\]

Where \(G_{i}\) represents the graph $i$ in the set, $S$, of the alloy phases present in the decomposition pathway, and \(G_{p}\) represents the final set of all unioned graphs (represented as a set of disconnected subgraphs)\cite{chapman2022efficient}. We note here that in our current construction of \(G_{p}\), the MPEA subgraphs is always represented by $i=0$.

\begin{figure}[h!]
    \centering
    \includegraphics[width=1.0\textwidth]{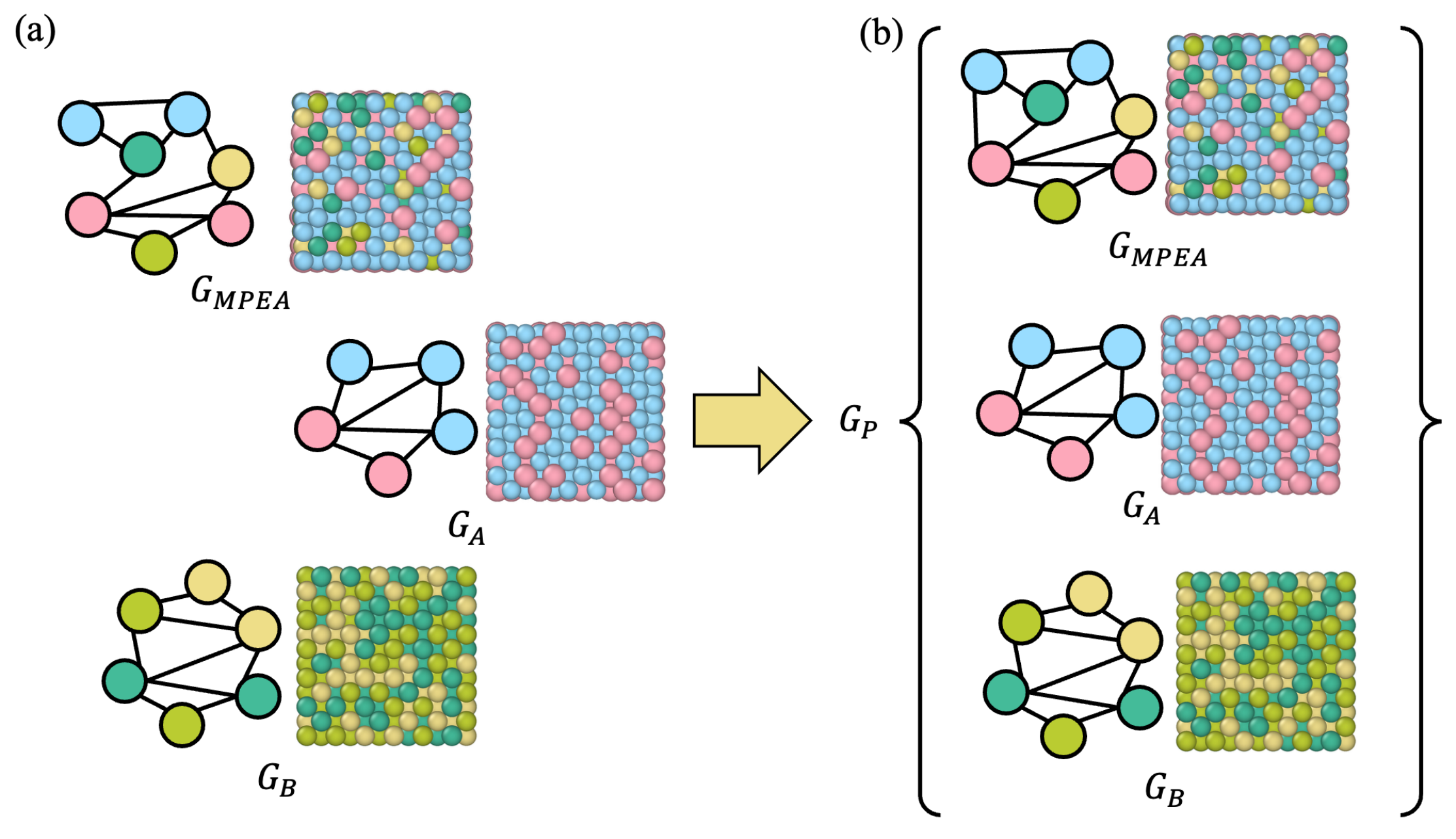}
    \caption{Example of the construction of \(G_P\). (a) The respective graphs are created for the MPEA \((G_{MPEA})\), phase alloy A \((G_{A})\), and phase alloy B \((G_{B})\). (b) All three graphs are then concatenated to form \(G_P\). Note that there are no edges connecting the graphs to each other, which is why \(G_P\) is considered a union of graphs rather than a super graph.}
    \label{fig:graph_creation}
\end{figure}

\textbf{Figure \ref{fig:graph_creation}} shows a schematic of the construction of \(G_{P}\). The model inputs are not individual MPEA graphs but rather \(G_P\), the graph representation that encodes the entire decomposition pathway. Consequently, the target property, \(y\), is defined as \(\bigtriangleup H_{mixing}\) computed via DFT.

The last layers are pooling layers that sum all the output features of \(G\) and output the embedding as a single vector with the same dimensions as the input vector. Since the convoluted values are passed individually through the NN, the output weights of the GNN are in the same order as the encoded features. This allows us to know which weight corresponds to which feature. For that reason, the weights are forced to be positive and sum to a positive scalar (\(\bigtriangleup H_{mixing}\)) before calculating the loss, and following the principles of feature-attribution this means the weights encode importance. 

\subsection*{GNN regression details}

In this work, we employ a GNN architecture, a batch size of 5, and 1500 training epochs. The batch size reflects hardware constraints and performance considerations, while the number of epochs was selected after preliminary trials with larger iteration counts, which indicated that 1500 epochs provided reliable convergence. The model was trained on an A100-80GB node. These and all other model parameters are summarized in \textbf{Table \ref{Table:2}}.

\begin{table}[h!]
    \centering
    \caption{ \label{Table:2} Model parameters used for the training and inference of the GNN regression model.}
    \begin{tabular}{|c c|}
        \hline 
        \rowcolor[HTML]{99DDFF}
        \textbf{Parameter} & \textbf{Value}\\
        \hline \hline
        Number of interaction layers & 6\\
        Radial Bessel basis cutoff  & 3.5 \AA \\
        Gaussian basis range & (-1,1) \\
        Number of channels & 100\\
        Batch size & 5\\
        Number of epochs & 1500\\
        Optimizer & AdamW \cite{adam2014method}\\
        Optimizer learning rate & 0.0001\\
        Training delta & 0.001\\
        Loss function & L1 (Mean Absolute Error) \cite{paszke2019pytorch} \\
        \hline
    \end{tabular}
\end{table}

A central goal of this framework is to evaluate the model's extrapolation capabilities, which we probe by separating the data according to distinct decomposition pathways. An MPEA can decompose into multiple alloy phases, and these pathways are categorized by the number of resulting phases. In this study, we focus on binary (two-phase) and ternary (three-phase) decomposition pathways, denoted as \(P_1\) and \(P_2\), respectively. The two pathway types differ not only in the number of product phases but also in the complexity of the compositional space they sample, making cross-pathway evaluation a meaningful test of generalization beyond the training distribution.

Accordingly, one model is trained exclusively on \(P_1\) data and another exclusively on \(P_2\) data. Each model is then tested on the pathway type it was not trained on, meaning the \(P_1\) model is evaluated on \(P_2\) data and vice versa. This cross-pathway evaluation scheme is designed to assess whether a model trained on simpler decomposition behavior can generalize to more complex phase spaces, and vice versa. In total, two models were trained for this study. The \(P_1\) dataset contains 877 graphs in total, of which 631 were used for training, 70 for validation, and 176 for testing. The \(P_2\) dataset contains 4442 graphs in total; however, due to hardware limitations, only a randomly selected subset of 2221 graphs was used, of which 1599 were used for training, 177 for validation, and 445 for testing, with the remaining 2221 graphs excluded from model development entirely.

Testing graphs were then identified via property binning in this latent space, ensuring that the test set spans the full range of the property distribution rather than clustering in densely sampled regions. The remaining graphs were subsequently divided into training and validation sets using Gaussian smearing, which promotes a smooth and representative coverage of the latent space across both subsets. The computational cost reflects these dataset sizes: training and testing the \(P_1\) model required approximately 2250 minutes and 86 minutes, respectively, while the \(P_2\) model required approximately 7980 minutes and 285 minutes, respectively, consistent with its larger training set.

\subsection*{GNN feature ranking details}

For the feature ranking analysis, we exploit the fact that each input descriptor is explicitly propagated through the GNN convolution layers and contributes in a traceable manner to the final prediction. A constant positive shift is applied to the target variable to to enforce non-negativity and then use a Positive Scalar Decoder through a softplus (or exponential) output head, so the loss can be computed under a strictly positive-valued objective. Following the final pooling operation, each feature in \(G_P\) is assigned a score value that reflects its relevance to the target property, and the sum of all scores reconstructs the predicted \(\bigtriangleup H_{mix}\). By aggregating the contributions within each descriptor class, we obtain a ranked ordering of the chemical and structural features, which is presented in the subsequent sections. 

We begin by defining the individual feature score  \(s_{i}^{(n)}\) for feature instance \(i\) of descriptor class \(n \in {1,2,3}\) as:

\begin{equation}
    s_{i}^{(n)} = \frac{d_i^{(n)}}{\hat{y}} 
\end{equation}

Where \(d_i^{(n)}\) is the raw decoder output and \(\hat{y}\) is the predicted \(\bigtriangleup H_{mix}\). \(n=1\) represents stoichiometric features, \(n=2\) 2nd-body interaction features, and \(n=3\) 3rd-body interaction features. Dividing by \(\hat{y}\) ensures that scores are normalized by the total predicted value, making them comparable across graphs. 

We then aggregate the scores for all structural instances of a given chemical motif \(C\) (e.g. Co, Fe-Ni, Co-Mo-Mo) within the MPEA structure:

\begin{equation}
    S_{C}^{(n)} = \sum_{i \in I_C} s_{i}^{(n)}
\end{equation}

Where \(I_C\) is the set of all structural instances of motif \(C\) in the MPEA graph. Note that only instances belonging to the MPEA structure are included. Finally, to enable comparison of feature contributions across Co concentration subsets, which may differ in the total number of feature instances per descriptor class, we normalize by the total aggregated score of descriptor class \(n\) within the MPEA:

\begin{equation}
    \tilde{S}_{C}^{(n)} = \frac{S_{C}^{(n)}}{\sum_{C} S_{C}^{(n)}}
\end{equation}

where the sum in the denominator runs over all chemical motifs \(C\) of class \(n\) present in the MPEA graph. By construction, \(\sum_{C} \tilde{S}_{C}^{(n)} = 1\), so \(\tilde{S}_{C}^{(n)}\) represents the fractional contribution of motif \(C\) to the total class-\(n\) score of the MPEA This is the quantity plotted in all subsequent figures, and we refer to it hereafter as the contribution of motif \(C\).
\section*{Results}
\subsection*{SPSS Latent Space Sampling}

To validate that the unsupervised GNN produces a physically meaningful latent space, we generated 500 MPEA structures for each of the unique stoichiometries studied, but varying atomic arrangements, and projected their graph embeddings onto the two leading latent coordinates \(X_1\) and \(X_2\) (\textbf{Figure \siref{3}}). Despite the composition remaining fixed, the latent space cleanly resolves structural diversity through differences in local chemical environments. This is quantified using the differential pair distribution function \(\Delta\rho_{i-[Mo-Mo]}(r)\), which measures how the local environment of a given atomic pair deviates from the Mo–Mo reference as a function of interatomic distance \(r\). Structures in the upper-left region of the latent space exhibit sharp, well-defined Fe–Ni and Ni–Ni correlation peaks, indicative of strong preferential short-range ordering between these species. 

In contrast, structures in the lower-right region show noticeably suppressed correlation features, consistent with more homogeneous distributions dominated by Mo–Mo interactions. The remaining corners capture intermediate ordering behaviors, confirming that position in the latent space continuously tracks with the degree and character of short-range chemical order. These results demonstrate that the encoder successfully disentangles structural variation even among configurations with the same composition, justifying its use as a sampling guide for selecting DFT training structures that span the relevant structural space.

\subsection*{DFT Calculations of Enthalpy of Mixing}
\begin{figure}[h!]
    \centering
    \includegraphics[width=1.0\textwidth]{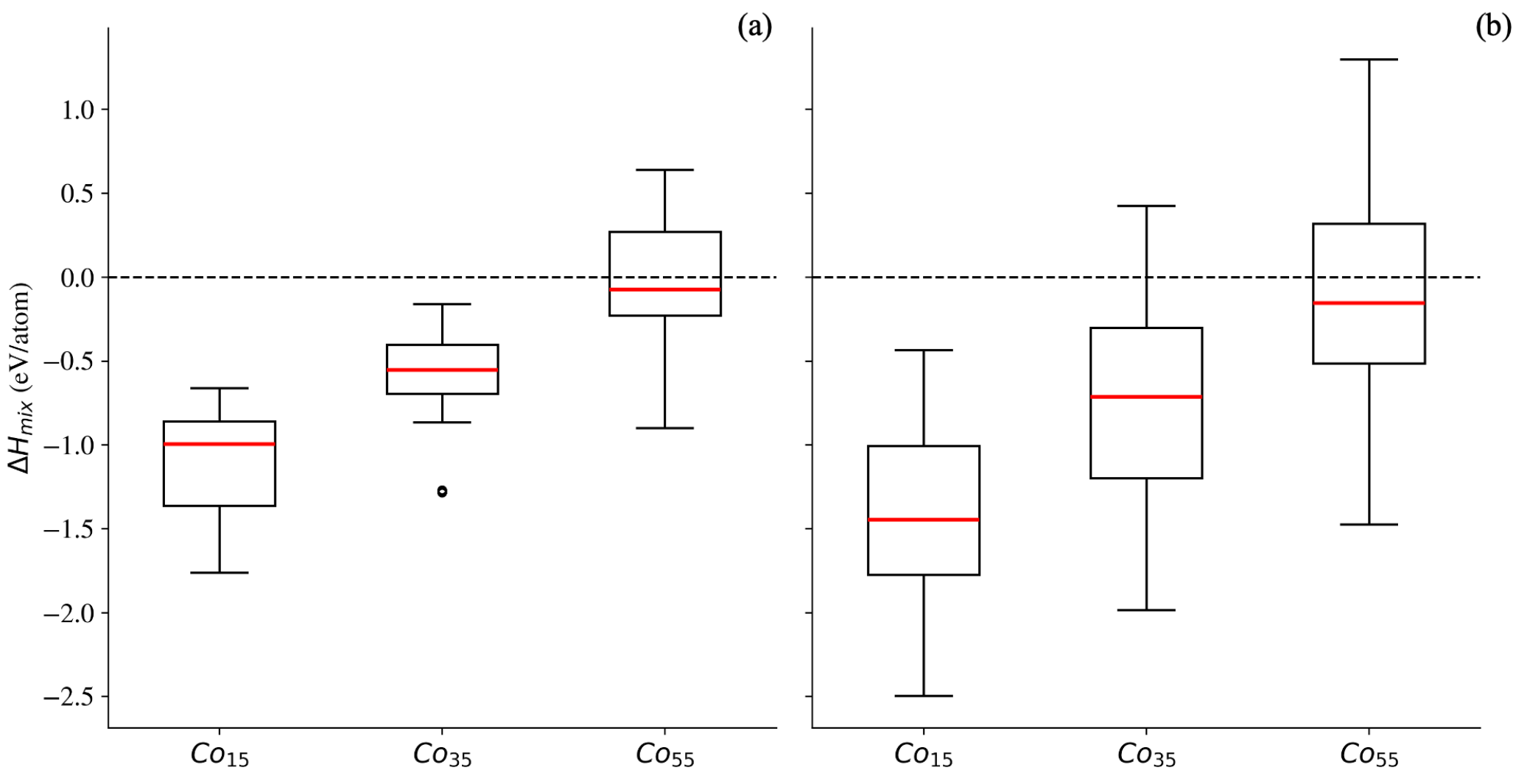}
    \caption{ \(\bigtriangleup H_{mix}\) distribution grouped by Co concentration for a) \(P_1\) and (b) \(P_2\) decomposition pathways. Dashed black line represents the crossover between energetically favorable vs unfavorable in the SPSS phase.}
    \label{fig:dft_results}
\end{figure}

DFT calculations provide useful preliminary insight into the thermodynamic stability of the CoMoFeNiCu system. \textbf{Figure \ref{fig:dft_results}} presents the distribution of the computed \(\bigtriangleup H_{mix}\) values for both \(P_1\) and \(P_2\) datasets, grouped by Co concentration. The results show a systematic shift of the \(\bigtriangleup H_{mix}\) distribution toward values above zero with increasing Co content, indicating a progressively greater thermodynamic driving force for decomposition. Accordingly, higher Co fractions appear to reduce the stability margin of the FCC single-phase solid solution, even though elevated Co fractions have been associated with increased catalytic activity \cite{saidi2021optimization}.

These findings are consistent with the available literature on this multicomponent alloy system. Xie et al. reported that CoMoFeNiCu can retain a single FCC solid-solution phase across a range of Co/Mo ratios comparable to those considered here, demonstrating that the system can remain in the FCC SPSS field despite compositional variation \cite{xie2019highly}. In the present study, the values of \(\bigtriangleup H_{mix}\)  generally remain near or below zero, supporting the persistence of the FCC SPSS state (\(92.13 \%\) for \(P_1\) pathways, \(86.36 \%\) for \(P_2\) pathways). However, the upward shift in the distributions with increasing Co content indicates that the thermodynamic resistance to phase separation decreases as Co is increased. More broadly, MPEA phase-stability studies have shown that changing the concentration of a single element can alter FCC/BCC phase competition by modifying valence electron concentration (VEC) and chemical driving forces, and in some composition ranges this can destabilize a nominally SPSS and promote secondary-phase formation \cite{kumar2025new}. This behavior reinforces the view that entropy alone does not guarantee SPSS, since composition-dependent enthalpic contributions can still drive decomposition.

\subsection*{GNN Model for Enthalpy of Mixing}

\begin{figure}[h!]
    \centering
        \includegraphics[width=1.0\textwidth]{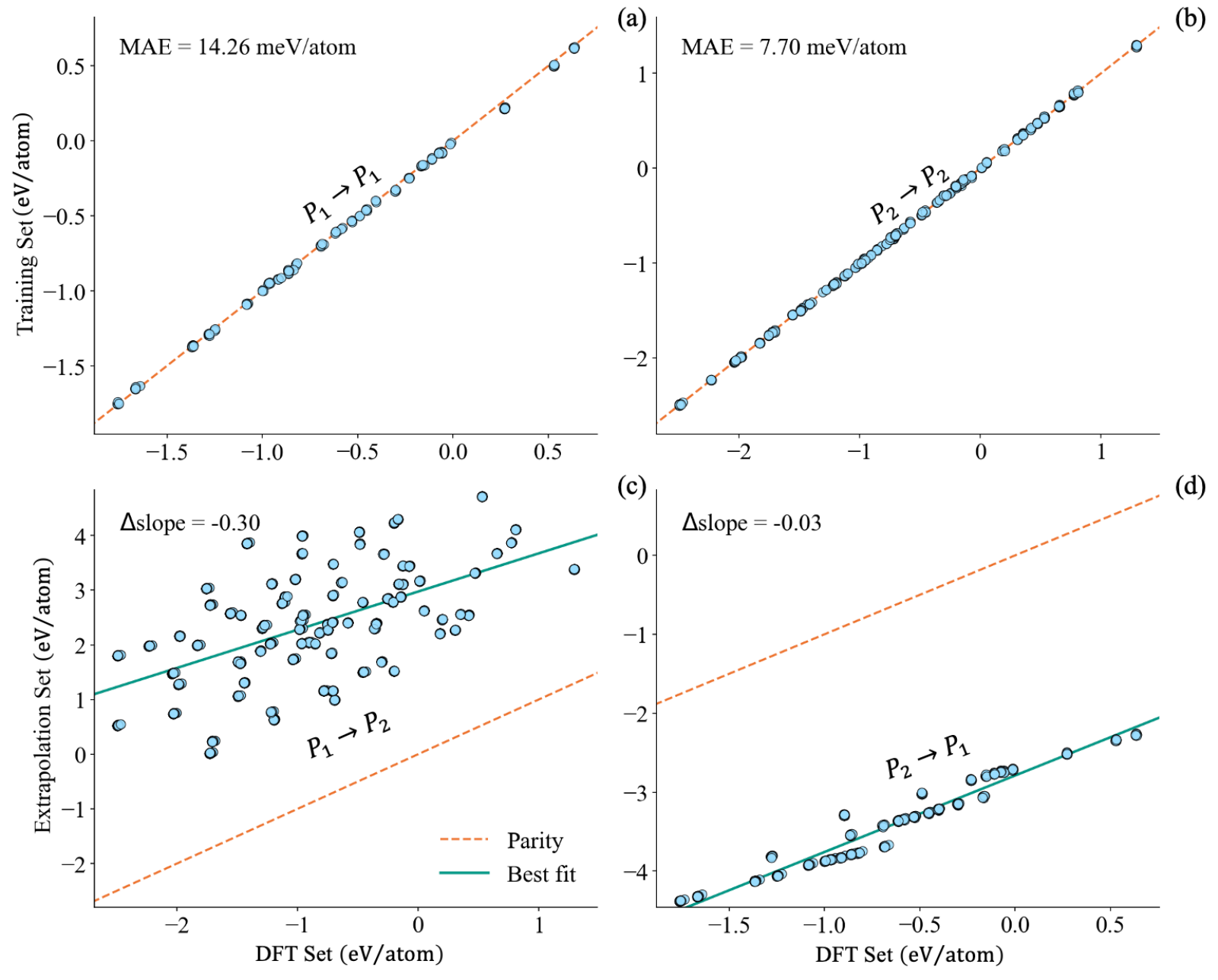}
    \caption{Parity plots of ML-predicted versus DFT-computed \(\Delta H_{\text{mix}}\) for both the \(P_1\) (left column) and \(P_2\) (right column) models, evaluated on their respective in-distribution test sets (top row) and cross-pathway extrapolation sets (bottom row). The dashed orange line represents ideal parity and the solid green line represents the best fit. In-distribution performance is strong for both models, with the \(P_1\) model achieving \(R^2 = 0.99\), MAE of \(14.26\frac{meV}{atom}\), and a hit rate of 100\%, and the \(P_2\) model achieving \(R^2 = 0.99\), MAE = \(7.70\frac{meV}{atom}\), and a hit rate of 99.8\%, with both MAE values at or below chemical accuracy. Cross-pathway extrapolation of the \(P_1\) model to \(P_2\) data (bottom left) exhibits a large systematic offset with \(\Delta\text{slope} = -0.301\), MAE (best fit) = \(665.43\frac{meV}{atom}\), and a hit rate of only 13.6\%, indicating limited transferability from binary to ternary decomposition pathways. By contrast, the \(P_2\) model (bottom right) captures the underlying trend of the \(P_1\) data considerably more reliably, with \(\Delta\text{slope} = -0.031\), MAE (best fit) = \(110.26\frac{meV}{atom}\), and a hit rate of 92.1\%, suggesting that training on the more compositionally diverse \(P_2\) dataset yields more transferable representations of the energy landscape.
}
    \label{fig:ml_results}
\end{figure}

We evaluated each model on its corresponding test set \(P_x \rightarrow P_x\), as illustrated by the parity plots in the top row of \textbf{Figure~\ref{fig:ml_results}}. These results demonstrate strong predictive performance for both models, with the \(P_1\) model (\textbf{Figure~\ref{fig:ml_results}(a)}) achieving \(R^2 = 0.99\) and MAE of \(14.26 \frac{\text{meV}}{\text{atom}}\), and the \(P_2\) model (\textbf{Figure~\ref{fig:ml_results}(b)}) achieving \(R^2 = 0.99\) and MAE of \(7.70 \frac{\text{meV}}{\text{atom}}\), with both values at or below chemical accuracy. It is worth noting that at this stage the models were deliberately allowed to overfit to their respective training sets. This choice was intentional as our primary goal here was not to assess extrapolation, but rather to evaluate the feature-ranking capabilities of the framework under conditions where the model has learned as much as possible about the data it was given.

By maximizing the model's fit to the training distribution, we ensure that the resulting feature scores reflect the full extent of what the model can learn about the relationship between structure and \(\Delta H_{mix}\). More broadly, this reflects a central goal of the framework: to demonstrate that physically meaningful features can be recovered from training and graph structure alone, without relying on hand-crafted descriptors or prior physical assumptions about which interactions should matter.

We also explored the model's extrapolative capabilities: the bottom row of \textbf{Figure~\ref{fig:ml_results}} presents the cross-pathway extrapolation results for both models \(P_x \rightarrow P_y\). The \(P_1\) model evaluated on \(P_2\) data (\textbf{Figure~\ref{fig:ml_results}(c)}) performs considerably worse in terms of numerical accuracy, with \(R^2 = 0.308\) and a best-fit MAE of \(665.43 \frac{\text{meV}}{\text{atom}}\) after a slope correction of \(\Delta\text{slope} = -0.301\), indicating a substantial systematic offset between predicted and reference values. This offset is consistent with the large energy scale difference between binary and ternary decomposition pathways, as the \(P_1\)-trained model has no exposure to the more complex compositional space of \(P_2\). The low hit rate of 13.6\% further confirms that the model struggles to correctly classify stability for pathways outside its training distribution. However, we emphasize that extrapolation is not the focus of the present study and is left for subsequent investigation

For the model trained on \(P_2\) data and evaluated on \(P_1\) data (\textbf{Figure~\ref{fig:ml_results}(d)}), a substantial systematic offset also persists, but the model captures the underlying trend considerably more strongly, reflected in an \(R^2 = 0.943\) and a best-fit MAE of \(110.26 \frac{\text{meV}}{\text{atom}}\) after a slope correction of \(\Delta\text{slope} = -0.031\), approaching chemical accuracy once the systematic shift is accounted for. Notably, the hit rate remains high at 92.1\%, indicating that although the numerical predictions carry a systematic offset, the model still classifies MPEA stability with considerable reliability.

\subsection*{GNN Feature Rankings}

\begin{figure}[h!]
    \centering
    \includegraphics[width=1.0\textwidth]{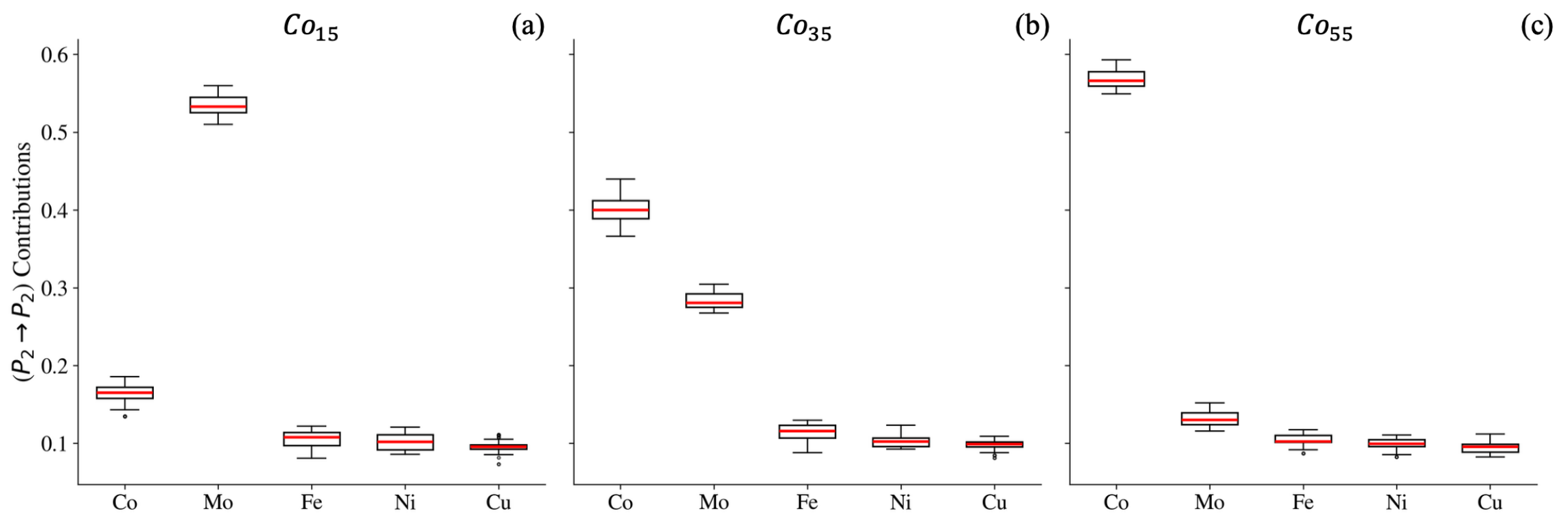}
    \caption{Normalized stoichiometric contributions for each element in the CoMoFeNiCu MPEA, grouped by Co concentration: (a) \textit{\(Co_{15}\)}, (b) \textit{\(Co_{35}\)}, and (c) \textit{\(Co_{55}\)}. The red line indicates the median and individual points denote outliers. A clear trend is observed with increasing Co content: Co's fractional contribution increases and Mo maintains its importance but is progressively overtaken as Co concentration increases. Fe, Ni, and Cu remain consistently low across all subsets, reflecting their fixed stoichiometric fractions and limited contribution to decomposition variability.}
    \label{fig:features_1}
\end{figure}

We summarize the feature-ranking results, focusing on the \(P_2\) model because it exhibits the strongest predictive performance. \textbf{Figure \ref{fig:features_1}} presents the normalized scores for the stoichiometric features, specifically the atomic species, grouped by Co concentration. At the lowest Co concentration, Mo receives the highest normalized score, while Co remains secondary, reflecting the Mo-rich stoichiometry of the \textit{\(Co_{15}\)} subset. As the Co fraction increases, this ordering reverses: Co's normalized contribution grows monotonically while Mo's decreases, and by \textit{\(Co_{55}\)} Co is clearly dominant. This composition-dependent redistribution of importance within the Co–Mo pair suggests that the model increasingly attributes the stability trend to the rising Co content and its effect on the local chemical environment. By contrast, Fe, Ni, and Cu remain consistently low in importance across all concentrations, which is consistent with their fixed stoichiometric fractions and their limited contribution to the variation in decomposition behavior. This behavior may also reflect Co-induced changes in valence electron concentration and atomic-size mismatch, both of which are established descriptors of MPEA phase stability and can shift the balance between FCC and BCC solid-solution formation as well as the thermodynamic cost of lattice distortion~\cite{guo2011effect,sheng2011phase,hu2024quantifying}.

Although the 2-body feature scores are much smaller in magnitude than the stoichiometric descriptors, they nevertheless display clear and consistent trends (\textbf{Figure \siref{4}}).At \textit{\(Co_{15}\)}, Mo–Mo and Co–Mo are the dominant pairwise interactions, with Co–Mo carrying the highest median contribution. As Co content increases to \textit{\(Co_{35}\)}, Co–Co emerges as co-dominant with Mo–Mo and Co–Mo, while at \textit{\(Co_{55}\)} Co–Co becomes the leading two-body interaction with a substantially higher median, followed by Co–Ni and Fe–Co. This progression indicates that pairwise bonding preferences within the variable local compositions carry the largest local relevance for \(\Delta H_{mix}\) even when global stoichiometry dominates the overall prediction.  Thus, the two-body ranking reinforces the picture from the stoichiometric features, the Co–Mo compositional axis sets the principal thermodynamic behavior governing decomposition pathways in this MPEA.

\begin{figure}[h!]
    \centering
    \includegraphics[width=1.0\textwidth]{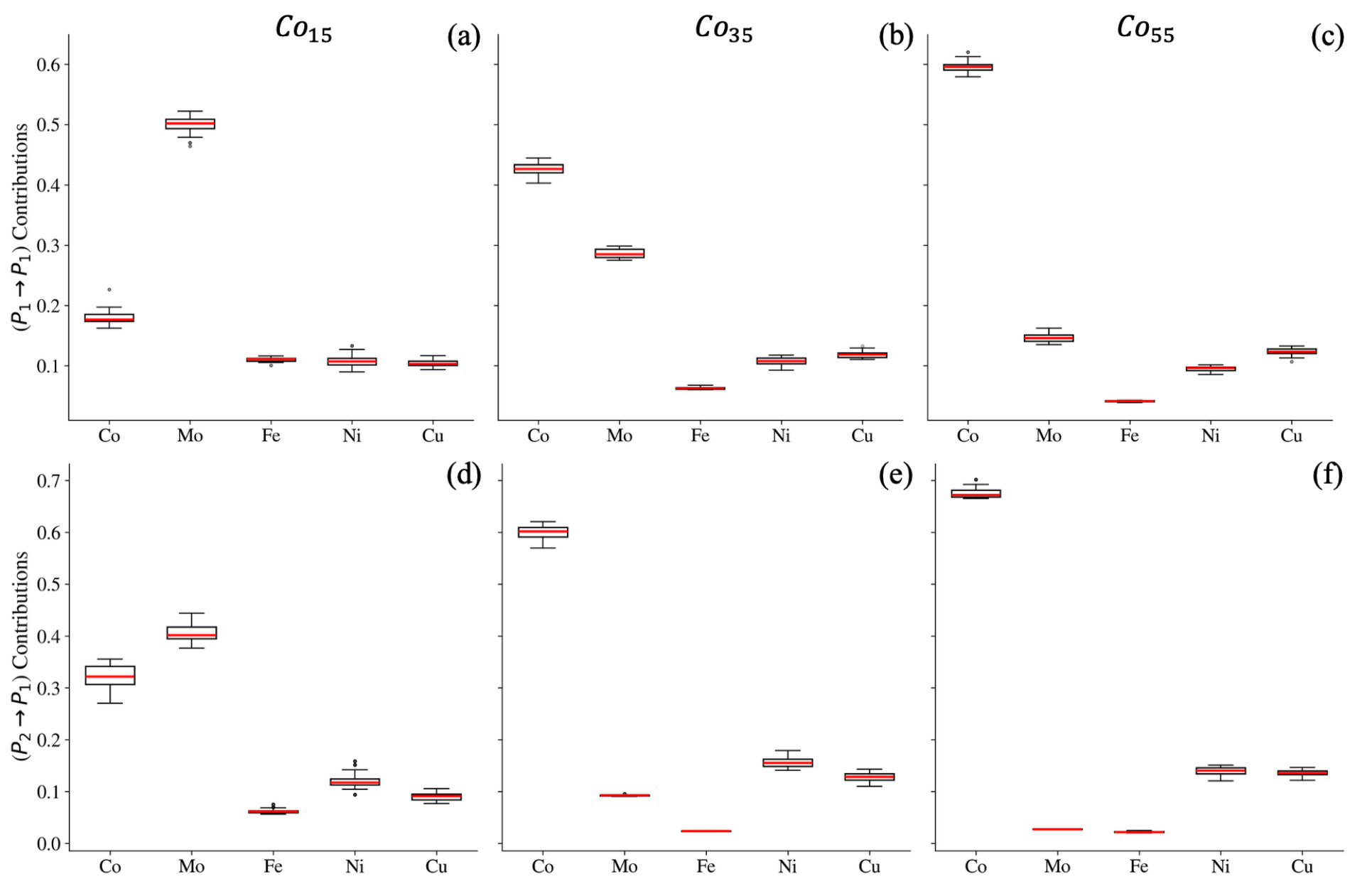}
    \caption{Normalized stoichiometric contributions per atomic species for \(P_1 \) pathways evaluated by both models, grouped by Co concentration: (a,b,c) \(P_1 \rightarrow P_1\ \), the binary-trained model evaluated on its own test set; (d,e,f) \(P_2 \rightarrow P_1\), the ternary-trained model evaluated on \(P_1\) pathways. Despite the \(P_2\) model being trained exclusively on ternary pathways and exhibiting poor predictive performance on binary structures, its feature attribution patterns closely mirror those of \(P_1 \rightarrow P_1\) across all three concentration subsets: Mo dominates at \textit{\(Co_{15}\)}, the Co–Mo balance shifts at \textit{\(Co_{35}\)} with Co overtaking Mo, and Co becomes overwhelmingly dominant by \textit{\(Co_{55}\)} while Fe collapses to near zero.}
    \label{fig:binary_results}
\end{figure}

Following the model's architecture, message passing on the line graph propagates angular information back into bond representations before each atom-level update, meaning not only pairwise bond contributions are captured but also higher-order local-environment effects associated with angular coordination or 3-body features (\textbf{Figure \siref{5}}), which are important in multi-principal-element alloys \cite{byggmastar2021modeling}. This also provides a natural explanation for why 3-body features are ranked on average higher than 2-body features; the 3-body angular environments are carried into the representation at every propagation step, making them a more expressive and information-rich descriptor than 2-body interactions. the three-body features are dominated throughout by mixed Mo–Co angular environments rather than same-species triplets. At \textit{\(Co_{15}\)}, Mo–Mo–Cu receives the highest median score, with Mo–Co–Co and Co–Co–Mo also elevated, while all other interactions remain comparatively low. As Co content increases to \textit{\(Co_{35}\)}, Mo–Co–Co and Co–Co–Mo emerge as clearly co-dominant, both showing substantially higher and similarly distributed score ranges compared to all other triplet types. By \textit{\(Co_{55}\)}, these two interactions remain the leading features, maintaining similarly high median scores while the rest of the distribution is suppressed. The consistent prominence of Mo–Co–Co and Co–Co–Mo across all three concentrations reinforces the conclusion from the stoichiometric and two-body analyses: the Co–Mo compositional axis is the primary driver of \(\Delta H_{mix}\) in this system, and the three-body encoding captures this relationship through the angular coordination environment between Co and Mo atoms, enriching the bond representation in a way that is geometrically consistent with the pairwise chemistry and stoichiometry.

\begin{figure}[h!]
    \centering
    \includegraphics[width=1.0\textwidth]{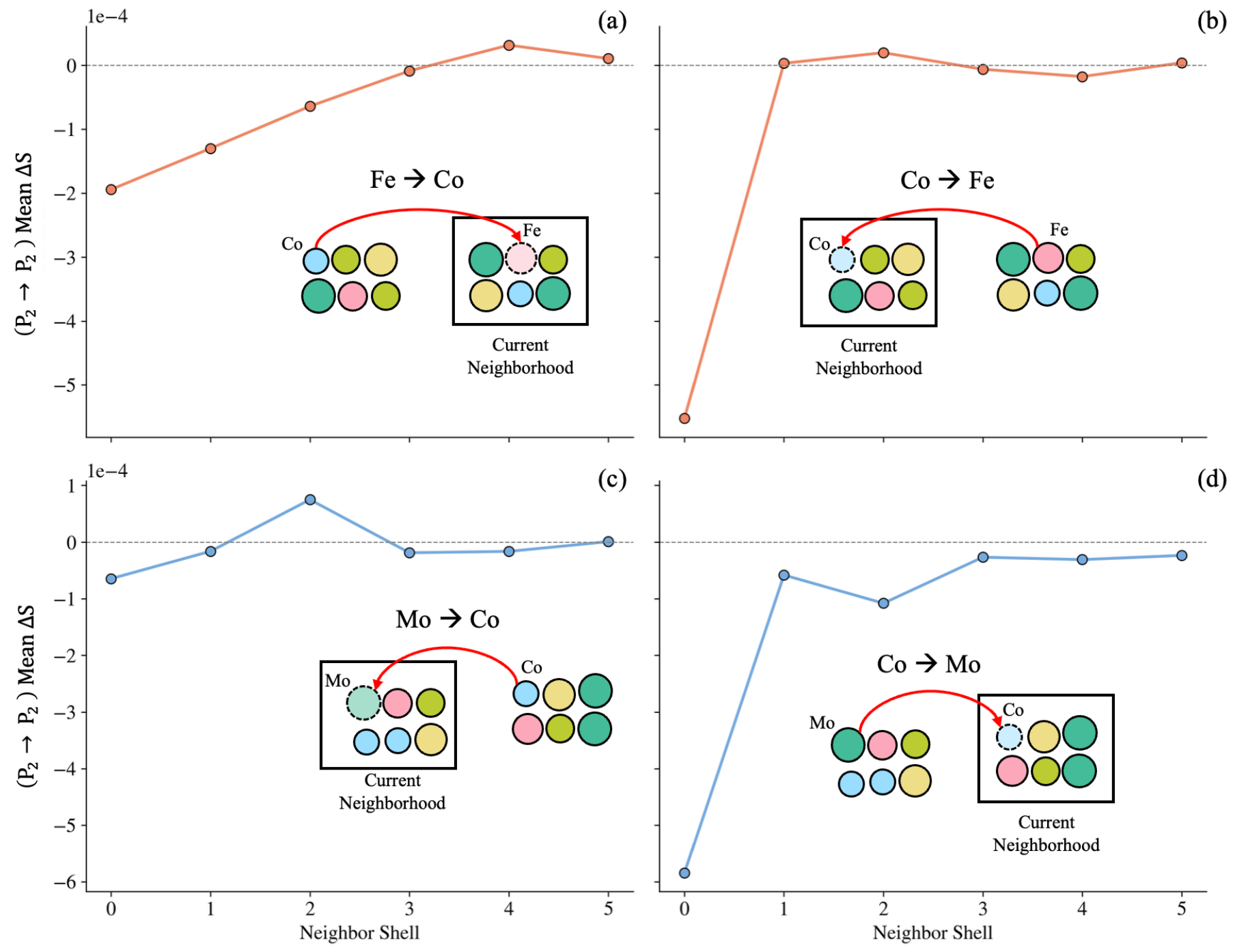}
    \caption{Mean score change (\(\Delta\text{S}\)) per neighbor shell following a single atomic swap, for two substitution pairs in the MPEA. For each plot, shell 0 represents only the swapped atom. The top row (a and b) corresponds to a high scoring Co \(\leftrightarrow\) low scoring Fe swap in a \textit{\(Co_{55}\)} structure, and the bottom row (c-d) to a high scoring Co \(\leftrightarrow\)  low scoring Mo swap in a \textit{\(Co_{35}\)} structure. Left and right columns show the two directions of each swap. The dashed line marks zero change. In both systems the largest perturbation is localised at shell 0, decaying toward zero within the first two to three shells, consistent with the short-range nature of the GNN's message-passing scheme.}
    \label{fig:shell_decay}
\end{figure}

To assess the generalization capabilities of the learned feature representations, we evaluated the stoichiometric score distributions for both models on \(P_1\) decomposition pathways. \textbf{Figure \ref{fig:binary_results} (a,b,c)} shows results for the \(P_1\)-trained model evaluated on its own test set, while \textbf{Figure \ref{fig:binary_results} (d,e,f)} shows the \(P_2\)-trained model evaluated on the same binary pathways. Remarkably, both figures display similar trends: Mo dominates at \textit{\(Co_{15}\)}, Co overtakes Mo as the leading species at \textit{\(Co_{35}\)}, and Co becomes overwhelmingly dominant at \textit{\(Co_{55}\)}, with Fe collapsing to near zero and Ni and Cu maintaining stable intermediate contributions throughout. This agreement is notable given that the \(P_2\) model was never trained on binary pathways and exhibits poor predictive performance in this extrapolation regime (as shown in \textbf{Figure~\ref{fig:ml_results}}). Despite the large systematic offset in predicted mixing enthalpies, the model's internal feature attribution remains physically consistent with what the binary-trained model recovers on the same structures.
\begin{figure}[h!]
    \centering
    \includegraphics[width=1.0\textwidth]{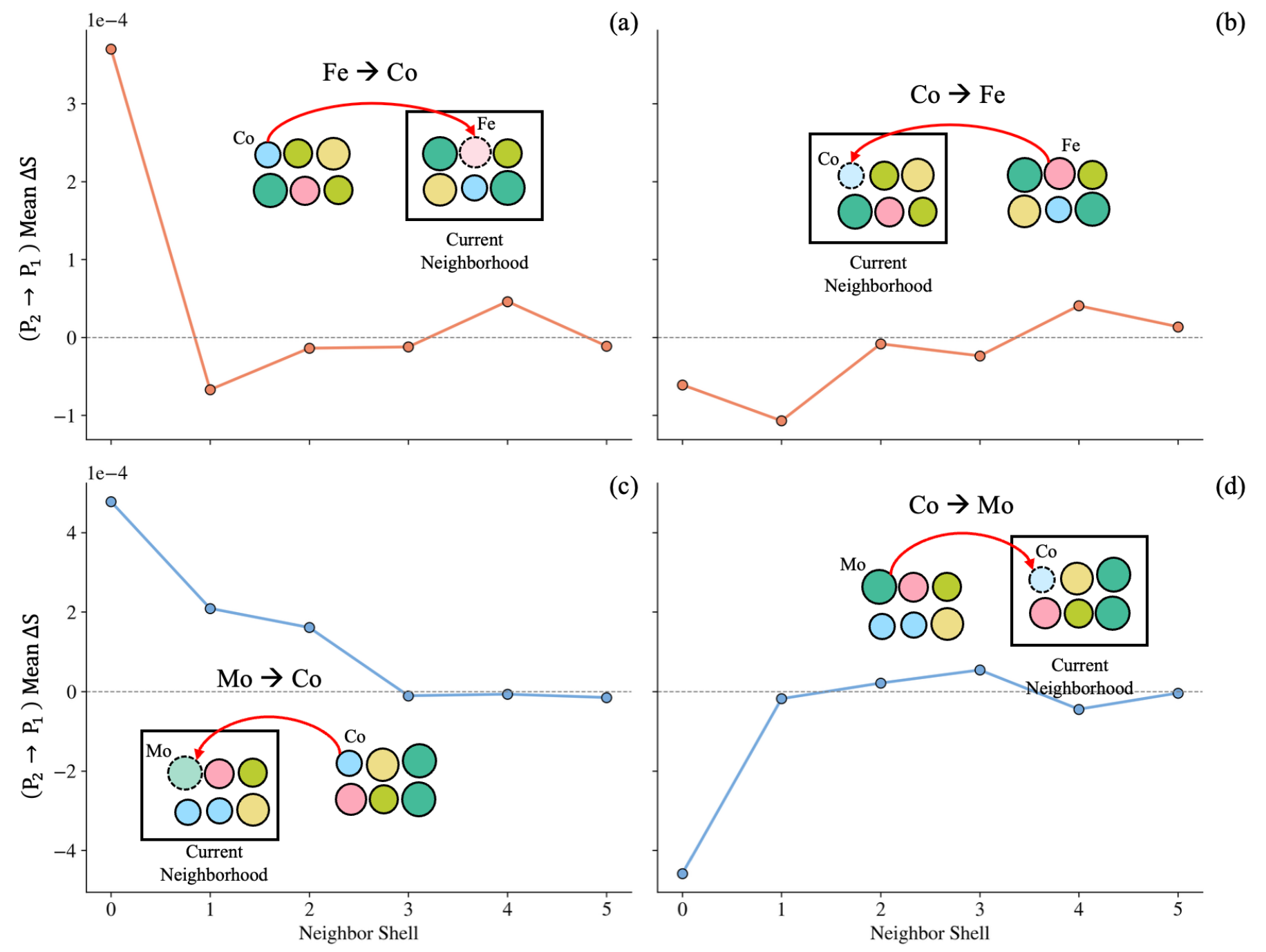}
    \caption{\(\Delta\text{S}\) per neighbor shell following a single atomic swap, for two substitution pairs in the MPEA for two \(P_1\) pathways tested on \(P_2\)-models. The top row (a and b) corresponds to a high scoring Co \(\leftrightarrow\) low scoring Fe swap in a \textit{\(Co_{55}\)} structure, and the bottom row (c-d) to a high scoring Co \(\leftrightarrow\)  low scoring Mo swap in a \textit{\(Co_{35}\)} structure. Left and right columns show the two directions of each swap. In both systems the largest perturbation is mostly localized at shell 0, decaying toward zero within the first two to three shells, consistent with the results from the \(P_2\) pathways.}
    \label{fig:shell_decay_binary}
\end{figure}

To further study the sensitivity of the model to changes in the local atomic environment, we selected two MPEA structures belonging to a $P_2$ pathway, one Co rich ($\textit{Co}_{55}$) and one with equal parts of Co/Mo ($\textit{Co}_{35}$), and exchanged the positions of a high-scoring Co atom with one low scoring Fe and one low scoring Mo, respectively. The same model was subsequently applied to the perturbed configurations (same decomposition pathway) to re-rank the features. \textbf{Figure \ref{fig:shell_decay}} further clarifies this effect by reporting the average change in score as a function of neighbor-shell distance from the swapped atom, for both the Co$\rightarrow$Fe / Fe$\rightarrow$Co exchange in the Co-rich structure (a, b) and the Co$\rightarrow$Mo / Mo$\rightarrow$Co exchange in the Co/Mo-balanced structure (c, d). 

In all four cases the score change is largest at shell 0, which corresponds only to the swapped atom, and remains elevated in the nearest shells. It then decreases progressively with increasing distance, consistent with the finite spatial range of message passing in the graph convolution. Because the model uses a cutoff of 3.5 \AA, corresponding approximately to the second neighbor shell, the strongest response is expected within the local neighborhood where atomic features are directly aggregated and then redistributed through successive convolution steps. In both cases, the score drops precisely at the swapped atomic site. This is expected: the sites previously occupied by Co atoms carried high scores due to Co's dominant role in governing \(\Delta H_{mix}\), and replacing Co with a lower-importance species causes the attribution at those sites to decrease accordingly. The reverse is also observed: sites that were originally occupied by the lower-scoring species and were subsequently replaced by Co show an increase in importance, consistent with Co's higher compositional relevance being transferred to those positions after the swap.

Despite the substantial local rearrangement in atomic scores, the predicted decomposition enthalpy of each structure was only marginally affected by the swap: for the Co-rich structure the DFT-calculated enthalpy shifted from 0.197 to 0.198~$\frac{eV}{atom}$ (predicted: 0.208$\rightarrow$0.186~$\frac{eV}{atom}$), and for the Co/Mo-balanced structure from -0.707 to -0.707~$\frac{eV}{atom}$ (predicted: $-0.707\rightarrow-0.69$3~$\frac{eV}{atom}$).We validated this behavior on binary decomposition pathways, as shown in \textbf{Figure~\ref{fig:shell_decay_binary}}. The results are qualitatively consistent with the ternary case: the largest score changes occur at shell 0 and decay rapidly with neighbor distance in all four swap configurations. Notably, swaps that introduce a higher-ranked species into the local neighborhood (Fe$\rightarrow$Co \textit{\(Co_{55}\)}, Mo$\rightarrow$Co in \textit{\(Co_{35}\)}) produce positive \(\Delta\) at shell 0, while the reverse swaps (Co$\rightarrow$Fe, Co$\rightarrow$Mo) produce negative values at shell 0, reflecting the reduction in local importance when a high-scoring atom is removed. Overall, these results indicate that while the model's intermediate, atom-level representations are locally sensitive to single-site substitutions, the aggregated, structure-level prediction remains consistent. This means the model is able to capture global properties while at the same time retaining atom-specific perturbation details.

\section*{Discussion}

The present results indicate that the GNN's feature rankings provide a coherent and physically interpretable picture of how local atomic environment descriptors govern the model's predictions across the CoMoFeNiCu system, with the Co–Mo sublattice emerging as the dominant compositional axis the model relies upon. As shown in \textbf{Figure~\ref{fig:features_1}}, as Co concentration increases the relative importance of Co grows while Mo remains substantial, with Co overtaking Mo as the leading species at \(Co_{35}\) and becoming overwhelmingly dominant by \(Co_{55}\). This suggests that the feature-importance landscape is not controlled by a single element in isolation, but by the changing interaction between Co and Mo. The weak contribution of Fe, Ni, and Cu across all compositions is also expected, since their stoichiometric fractions are fixed and they therefore contribute less to the observed variation captured by the model.

The pairwise and three-body interaction analysis further supports this interpretation. As shown in \textbf{Figure~\siref{4}}, the same dominant chemical motif appears repeatedly across concentrations: Co–Co and Mo–Mo interactions rank highest, with Co–Mo mixed pairs also prominent. This indicates that the model is sensitive to how the changing Co content reshapes local bonding environments rather than only to average composition. More notably, the three-body terms (\textbf{Figure~\siref{5}}) are more influential than the two-body terms, which implies that higher-order local-environment effects are important to the model's representation. This is physically reasonable for a chemically complex alloy, where local coordination, angular dependence, and short-range order can significantly affect feature relevance~\cite{mccarthy2023atomic}. The predominance of Mo–Co–Co and Co–Co–Mo triplets across all three concentration subsets supports the conclusion that the Co–Mo chemical subspace controls not only the model's average feature attribution but also its local structural sensitivity.

A key observation is that these rankings exhibit a degree of generalizability beyond the training domain.  Despite the \(P_2\) model being trained exclusively on ternary pathways and showing poor predictive performance on binary structures \(R^2 = 0.26\), its feature attribution patterns are qualitatively similar to those of the \(P_1\) model (\textbf{Figure \siref{6}} and \textbf{Figure \siref{7}}). This suggests that the Co–Mo compositional signal is sufficiently strong and universal that it is captured by both models regardless of training domain, and that feature rankings can carry meaningful physical interpretability even when absolute predictions fail. In other words, the GNN encodes chemically relevant structure–property relationships at the feature level that persist across training distributions, pointing to the possible robustness of the ranking system as a descriptor of compositional importance beyond the boundaries of the training set and hence a feature worth investigating further.

An additional consideration that may help explain the stability and interpretability of these feature rankings is the use of a positive scalar decoder in the model's output. By constraining the decoder weights to strictly positive values, we restrict the learned representation to a smaller subspace. This means that the decoder acts as an implicit regularizer on the feature attributions, encouraging the model to concentrate importance on a smaller subset of features that reliably drive the output in one direction, rather than having features cancel each other out through opposing contributions. We plan to investigate this hypothesis more directly in future work.

These trends can be interpreted in the context of established MPEA design concepts. Increasing Co content likely modifies the effective VEC and atomic-size mismatch, both of which are known to influence local bonding environments and phase competition in multicomponent alloys~\cite{zhang2022effect}. In that sense, the model's redistribution of feature importance with composition is not unexpected: it is learning descriptors that are already known to be physically relevant in governing local atomic interactions in such systems. The fact that the model's interpretation shifts coherently with composition, rather than remaining static, suggests that the learned representation captures a genuine and physically meaningful sensitivity to the underlying chemistry rather than a statistical artifact.

Looking ahead, this framework can be extended to a broader set of MPEA structures to test whether these trends generalize across different chemistries and phase fields. Such an expansion would allow a more systematic study of compositional effects on feature attribution and could provide a basis for symbolic regression using the GNN-derived scores, with the goal of extracting interpretable analytical relationships for MPEA design.

\section*{Data availability}
The code and metadata needed to reproduce this work can be found at https://github.com/Materials-Informatics-Laboratory/Catalyst. Metadata needed to reproduce this work can be found at https://github.com/Materials-Informatics-Laboratory/Public\_database.

\section*{Acknowledgments}
MT and JC acknowledges support from the Department of Mechanical Engineering and the College of Engineering at Boston University. JC acknowledges support from NSF ACCESS grant MAT230025.
\printbibliography

\end{document}